\documentclass[aps,prl,twocolumn,10pt,superscriptaddress,colorlinks=true,allcolors=blue,amsmath]{revtex4-2}
\pdfoutput=1

\usepackage{microtype}
\usepackage{silence}
\usepackage{comment}

\usepackage[utf8]{inputenc}
\usepackage[breaklinks]{hyperref}
\usepackage[ 
    paperwidth=210mm,
    paperheight=276mm,
    textwidth=180mm,
    textheight=243.5mm,
    columnsep=4mm, 
]{geometry}
\usepackage[british]{babel}
\usepackage[babel=true]{csquotes}

\usepackage{graphicx}
\graphicspath{{Figures/}}
\usepackage{xcolor}
\usepackage{placeins}

\usepackage{amsmath}
\usepackage{siunitx}
\DeclareSIUnit{\fps}{ \translate{frames per second} }
\usepackage{textgreek}
\usepackage{textcomp}
\usepackage{lmodern}
\usepackage{tabularx}

\setcitestyle{super} 

\usepackage{titlesec}
\titleformat*{\section}{\raggedright\bfseries\sffamily\large}
\titlespacing*{\section}{0em}{1em}{0.5em}
\titleformat{\subsection}[runin]{\raggedright\bfseries}{}{}{}[.]
\titlespacing*{\subsection}{0em}{1em}{0.5em}
\usepackage{lettrine}

\usepackage{caption}
\let\vec\mathbf

\usepackage{mathtools}
\DeclarePairedDelimiter\abs{\lvert}{\rvert}%
\DeclarePairedDelimiter\norm{\lVert}{\rVert}%

\makeatletter
\let\oldabs\abs 
\def\abs{\@ifstar{\oldabs}{\oldabs*}}
\let\oldnorm\norm 
\def\norm{\@ifstar{\oldnorm}{\oldnorm*}}
\makeatother

\let\oldfrac\frac
\makeatletter
\newcommand{\groupit}[1]{(#1)}
\newcommand{\nogroupit}[1]{#1}
\renewcommand{\frac}[2]{%
  \setbox\z@\hbox{$#1$}
  \setbox\tw@\hbox{$#2$}
  \ifdim\wd\z@>1em \let\groupornot@i\groupit\else\let\groupornot@i\nogroupit\fi
  \ifdim\wd\tw@>1em \let\groupornot@ii\groupit\else\let\groupornot@ii\nogroupit\fi
  \mathchoice
    {\oldfrac{#1}{#2}}
    {\groupornot@i{#1}/\groupornot@ii{#2}}
    {\groupornot@i{#1}/\groupornot@ii{#2}}
    {\groupornot@i{#1}/\groupornot@ii{#2}}
}
\makeatother

\newcommand{\iu}{{\mathrm{i}}}

\newcommand{\exoref}[2]{\hyperref[#1]{\ref*{#1}#2}}
\newcommand{\lref}[2]{\hyperref[#1]{#2}}

\newcommand{\MPINAT}{Max Planck Institute for Multidisciplinary Sciences, 37077 Göttingen, Germany}
\newcommand{\UGOE}{4th Physical Institute -- Solids and Nanostructures, University of Göttingen, 37077 Göttingen, Germany}
\begin{document}


\title{\sffamily\Large
Attosecond electron microscopy by free-electron homodyne detection %
}
\author{John H. Gaida}
\author{Hugo Lourenço-Martins}
\author{Murat~Sivis}
\author{Thomas~Rittmann}
\author{Armin~Feist}
\affiliation{\MPINAT}
\affiliation{\UGOE}
\author{F. Javier García de Abajo}
\affiliation{ICFO-Institut de Ciencies Fotoniques, The Barcelona Institute of Science and Technology, 308860 Castelldefels (Barcelona), Spain}
\affiliation{ICREA-Instituci\'o Catalana de Recerca i Estudis Avan\c cats, 08010 Barcelona, Spain}
\author{Claus Ropers}
\email[Corresponding author: ]{claus.ropers@mpinat.mpg.de}
\affiliation{\MPINAT}
\affiliation{\UGOE}

\date{\today}

\begin{abstract}
Time-resolved electron microscopy aims at tracking nanoscale excitations and dynamic states of matter with a temporal resolution ultimately reaching the attosecond regime. Periodically time-varying fields in an illuminated specimen cause free-electron inelastic scattering, which enables the spectroscopic imaging of near-field intensities. However, access to the evolution of nanoscale fields and structures within the light cycle requires a sensitivity to the optical phase.
Here, we introduce Free-Electron Homodyne Detection (FREHD) as a universally applicable approach to electron microscopy of phase-resolved optical responses at high spatiotemporal resolution. In this scheme, a phase-controlled reference interaction serves as the local oscillator to extract arbitrary sample-induced modulations of a free-electron wave function. We demonstrate this principle through the phase-resolved imaging of plasmonic fields with few-nanometer spatial and sub-cycle temporal resolutions. Due to its sensitivity to both phase- and amplitude-modulated electron beams, FREHD measurements will be able to detect and amplify weak signals stemming from a wide variety of microscopic origins, including linear and nonlinear optical polarizations, atomic and molecular resonances and attosecond-modulated structure factors. 

\end{abstract}

\maketitle

\begin{figure}[tb]
\centering
\includegraphics[]{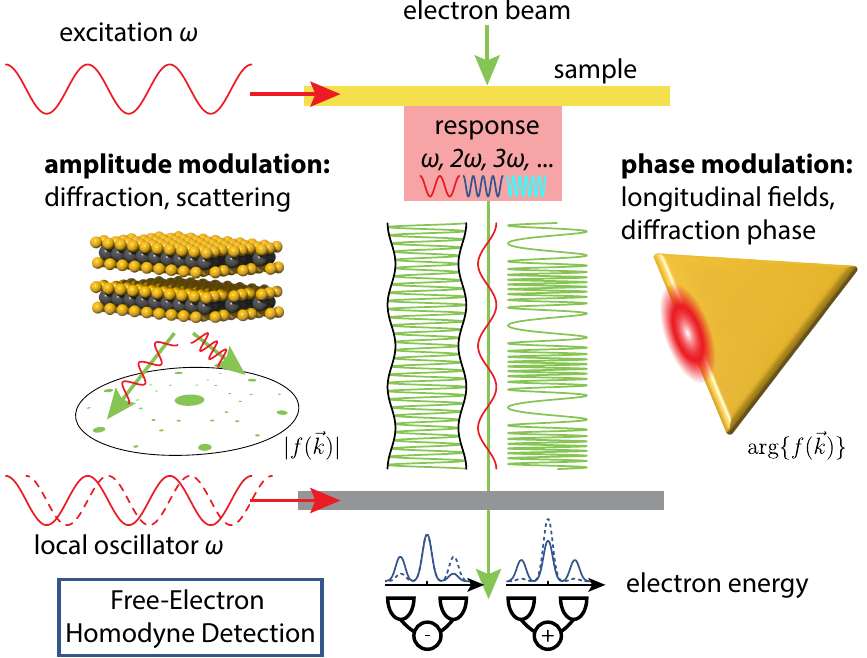}
\caption{
\textbf{Measurement of arbitrary amplitude and phase responses by Free-Electron Homodyne Detection.} 
Optical excitation of an investigated sample at frequency $\omega$, inducing a response at the fundamental driving frequency and its harmonics. A transmitted or diffracted electron beam experiences a modulation in its amplitude and/or phase that traces the response. For example, a modulation of the magnitude of the structure factor $f(\protect\overrightarrow{k})$ of a material leads to an amplitude modulation of a diffracted electron wave function (left), while localized optical fields and polarizations typically result in a phase modulation (right). A second interaction with a local oscillator, serving as a reference or mixer with variable phase, yields antisymmetric and symmetric signals, respectively, in the final electron kinetic energy spectrum, which is measured in this homodyne detection scheme.
}\label{fig:homodyne-principle}
\end{figure} 

The desire to map the structure and dynamic evolution of materials on their intrinsic spatiotemporal scales of {\AA}ngströms and attoseconds has been a major driving force behind methodological developments in condensed matter science. While structural information is available from X-ray~\cite{Chapman2011, Miao2015, Zayko2021} and electron~\cite{Scott2012, Yang2017, Yip2020} imaging and diffraction, temporal resolution down to the attosecond regime is provided by a growing suite of methods in optical spectroscopy~\cite{Corkum2007, Krausz2009, Calegari2016}. Ultrafast electron microscopy combines the strengths of optical techniques with nanoscale spatial resolution~\cite{King2005, Zewail2010, Piazza2013, Feist2017, Houdellier2018, Zhu2020} for imaging non-equilibrium phenomena involving structural phase transformations~\cite{Danz2021}, strain wave and polariton propagation~\cite{McKenna2017, Feist2018a, Kurman2021}, local spin dynamics~\cite{RubianodaSilva2018} and many other phenomena~\cite{Zandi2020}. In this approach, electrons are synchronised with optical excitations driven by femtosecond laser pulses, commonly used to trigger dynamical processes in solids, nanostructures and molecules. The characterisation of the associated microscopic couplings and correlations with ultimate spatiotemporal resolution will facilitate future applications in materials synthesis, energy conversion and light harvesting.

The response of a material to an optical stimulus includes both linear and nonlinear contributions~\cite{Ghimire2019, Dombi2020}, which intrinsically involve a temporal evolution below the optical period~\cite{Herink2012, Langer2016, Higuchi2017, Spektor2017, Reimann2018}. To probe such dynamics, pulses with a sub-cycle temporal structure are employed~\cite{Cavalieri2007, Schultze2010}. For example, high-harmonic generation facilitates the creation of attosecond extreme ultraviolet pulses~\cite{Sansone2006} or pulse trains~\cite{Antoine1996, Hentschel2001} for spectroscopy. In electron microscopy, the generation of energy sidebands by coherent inelastic electron scattering~\cite{Barwick2009, GarciadeAbajo2010, Park2010} allows sub-cycle bunching of the electron beam through dispersive propagation~\cite{Feist2015}, promising attosecond microscopy~\cite{Baum2007, Priebe2017, Morimoto2018, Kozak2018, Hassan2018, Ryabov2020} with high spatial resolution. Recent works demonstrated attosecond electron pulse trains within the stringent spatial constraints imposed by electron microscopes~\cite{Priebe2017, Schonenberger2019, Ryabov2020}, but nanometric attosecond imaging remains a major challenge.

In this Letter, we introduce Free-Electron Homodyne Detection (FREHD) as a universal scheme for attosecond electron microscopy of arbitrary periodic sample responses. The technique is based on a nanoscale readout of coherent amplitude or phase modulations of the free-electron wave function. Not requiring a density structuring of the beam, FREHD will enable the spatially resolved mapping of sub-cycle optical, electronic or structural dynamics. We experimentally demonstrate the capabilities of this approach in the phase-resolved imaging of plasmonic near fields at the few-nanometre scale. 

The method presented here transfers the notion of homodyne detection in optics and radio frequency technology to electron beams, with conceptual similarities but also notable additional features. In optics, homodyne detection is frequently used to characterise states of light in relation to a local oscillator~\cite{Mandel1995, Schleich2001}. Specifically, the signal to be probed is brought to interference with a reference wave derived from the same source (the "local oscillator") by mixing both at a beam splitter. A key advantage of this scheme is a sensitivity to the relative phase between the signal and the reference. Moreover, a coherent amplification of the signal of interest is accomplished using a sufficiently strong local oscillator, although shot-noise constraints for weak signals are generally known to remain in place~\cite{Yuen1983}.

In analogy, Free-Electron Homodyne Detection facilitates the nanoscale probing of microscopic sample responses by an electron beam in a phase-resolved manner and with an added sensitivity to non-linear sub-cycle temporal evolution. Figure~\ref{fig:homodyne-principle} schematically illustrates the principle of the technique. We consider an optical excitation at a frequency $\omega$ that induces a response in an investigated specimen, generally containing contributions at both the fundamental driving frequency and its harmonics $2\,\omega, 3\,\omega$,~$\dots$
An electron beam transmitted through or diffracted by the sample experiences a modulation governed by the detailed electronic and structural response to the excitation. For example, a time-periodic variation of the magnitude of the structure factor in a crystalline specimen~\cite{Yakovlev2015} leads to an amplitude modulation of the electron wave function along its propagation direction (left part of Fig.~\ref{fig:homodyne-principle}). In contrast, phase oscillations of the structure factor, as well as localized optical near fields and nonlinear optical polarizations~\cite{Konecna2020} with vector components along the beam path, result in a longitudinal phase (i.e., momentum) modulation of the wave function by inelastic electron--light scattering (right part of Fig.~\ref{fig:homodyne-principle}), as leveraged in photon-induced near-field electron microscopy~\cite{Barwick2009, GarciadeAbajo2010, Park2010, Yurtsever2012a, Piazza2013, Feist2015} (PINEM).

Consequently, upon traversing a sample plane at velocity $v_\text{e}$, the electron wave function is modulated in its amplitude and/or phase, with complex parameters of the $n$-th harmonic modulation $a_n=\abs{a_n}\exp(\iu\varphi_{\text{a},n})$ (amplitude-modulation coefficient) and $g_n=\abs{g_n}\exp(\iu\varphi_{\text{g},n})$ (phase-modulation coefficient). In the wave function, pure amplitude or phase modulations correspond to prefactors of the type $[1+\abs{a_n}\sin(n\frac{\omega}{v_\text{e}} z + \varphi_{\text{a},n})]$ or 
$\exp[-2\iu\abs{g_n}\sin(n\frac{\omega}{v_\text{e}} z + \varphi_{\text{g},n})]$, respectively. Generally, combinations of both types of modulations at different harmonics will be possible, and likely occur in diffractive probing of strongly light-driven charge densities~\cite{Langer2016, Ghimire2019}. In the past, electron phase modulation was primarily considered, generally resulting in multiple higher-order sidebands in the electron energy spectrum~\cite{Barwick2009, Piazza2013, Piazza2015, Feist2015, Kfir2020, Henke2021, Shiloh2022, Auad2022}. Yet, in direct analogy to radio-frequency technology amplitude modulation is expected to produce a single pair of energy sidebands for each harmonic.

For illustration, we consider the sample responses $g$ and $a$ at the fundamental frequency only (dropping the index $n$). In the limit of weak interactions, these responses generate electron energy sidebands with populations $P_{\pm 1}~\propto \abs{g}^2,\abs{a}^2$ (for $|g|, |a| \ll 1$), separated from the initial beam energy by $\pm\hbar \omega$ (Refs.~\citenum{Barwick2009, GarciadeAbajo2010, Park2010, Feist2015}).
We are now interested in the quantitative and phase-resolved determination of these modulations, which constitute the attosecond sample response. This is achieved by inducing a subsequent coherent and phase-controllable light interaction with the electron beam, which serves as a local oscillator (i.e., a reference) for the electron wave function modulations. Either amplitude or phase modulation could be used as the local oscillator, but for simplicity and experimental practicality  (i.e., readily available through PINEM), we consider a phase-modulating local oscillator $g_\text{ref}$ with a controllable phase \(\varphi_\text{ref} = \text{arg}\{g_\text{ref}\}\).
The local oscillator leads to final-state interference in the electron energy sidebands, in distinctive manners for amplitude and phase modulations. Specifically, the symmetric~($S$) and antisymmetric~($A$) components of the first-order sideband populations directly encode the phase-modulation and amplitude-modulation response of the material, respectively. Approximated to first order in the strength of the reference, these independent FREHD signals become 
\begin{align}
 & S=P_{+1}+P_{-1}\propto \abs{g}\abs{g_\text{ref}}\cos(\varphi_\text{g}-\varphi_\text{ref}) + \text{const.} \\
& A=P_{+1}-P_{-1}\propto \abs{a}\abs{g_\text{ref}}\cos(\varphi_\text{a}-\varphi_\text{ref}) + \text{const.}
 \label{eq:Amp_Phase}
\end{align}
These complementary dependencies arise from the conjugate symmetric versus asymmetric complex amplitudes produced by the two different kinds of modulations (see also Extended Data Fig.~\ref{fig:quantum_phasespace}). A generalization of these expressions to arbitrary harmonic responses and references requires the application of free-electron quantum state reconstruction, as recently introduced~\cite{Priebe2017, Shi2020}.

\begin{figure}[tb]
\centering
\includegraphics[]{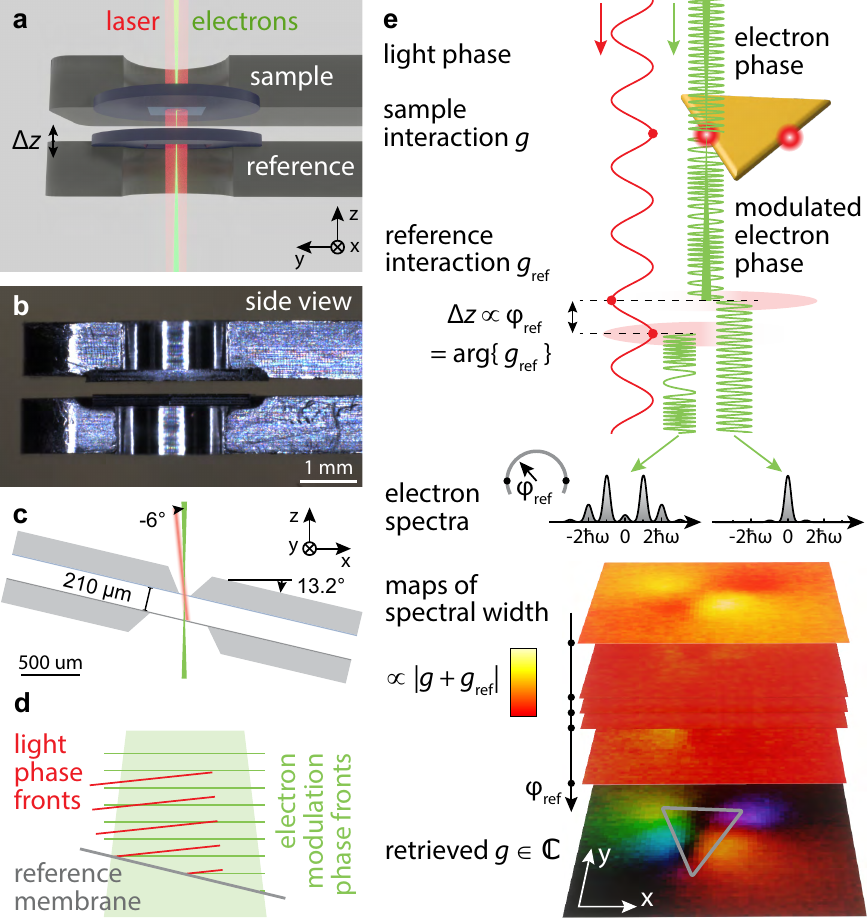}
\caption{
\textbf{
Implementation of free-electron homodyne detection at high spatial resolution
}
\textbf{a},~Schematic of focused electron beam and co-propagating laser beam passing the sample and reference. A dedicated dual-plane holder allows for an independent vertical displacement of the reference (here, a silicon membrane).
\textbf{b},~Side view of the piezo-controlled holder.
\textbf{c},~Geometry of the experimental configuration. Focused to a {5}-{nm} spot size at the sample, the electron beam has diverged to a beam diameter of about \SI{13}{\micro m} at the reference. To ensure electron--light phase matching, the sample is tilted by a predetermined angle. In this way, the phase fronts of the laser light and of the modulated electron wave function are matched at the reference membrane (see \textbf{d}). 
\textbf{e},~Stimulated inelastic scattering at an investigated nanostructure (here, a plasmonic triangle) leads to quantum-coherent electron-phase modulation. The displacement of the reference plane controls the phase of the reference interaction $\varphi_\text{ref}$ and the resulting final sideband populations. Bottom: The magnitude of the total phase modulation is measured by raster scanning, recording kinetic energy spectra at each in-plane position and for a set of reference phases. These interferograms yield a spatial map of the complex phase modulation induced by the sample.
}\label{fig:experiment-concept}
\end{figure}

From a fundamental viewpoint, the described approach exploits the quantum coherence of sequential interactions with free-electron beams, previously demonstrated in the context of Ramsey-type phase control~\cite{Echternkamp2016}, utilised for the quantitative measurement of attosecond electron pulses~\cite{Priebe2017}, and underlying recent demonstrations of cathodoluminescence interference from sequential scatterings~\cite{Taleb2023}. Analoguously, sequential interactions have been employed to create energy-filtered holograms of travelling surface plasmons~\cite{Madan2019}. Harnessing such phase-locked interactions in a universally applicable microscopy scheme to retrieve sub-cycle information requires both high spatial resolution and a controlled means of varying the phase of the local oscillator in a way that is independent of the sample under investigation. In essence, such a technique must combine free-electron quantum-state reconstruction, achieved only for collimated beams to date~\cite{Priebe2017}, with high spatial resolution.

In the following, we experimentally implement and demonstrate FREHD for resolving the plasmonic near-field evolution at a gold nanoprism with sub-cycle temporal (i.e., phase) resolution using a {5}-{nm} probe beam. The measurements are conducted at the Göttingen Ultrafast Transmission Electron Microscope (UTEM), which is equipped with a laser-triggered electron gun that generates photoelectron pulses for the femtosecond probing of structural, magnetic and optical excitations~\cite{Feist2017}.
Figure~\ref{fig:experiment-concept} illustrates the geometry employed in the experiment. As a key element, a custom-designed double specimen holder allows for the positioning of a reference membrane sample in proximity (here: \SI{210}{\micro m}) below\footnote{We note that placing the reference above the sample is of course equally possible. Generally, however, dispersive effects must be considered, which will differ in both scenarios} the specimen under investigation. Using a piezo actuator between the two arms of the double holder, the distance \(z\) between the sample and the reference membrane can be precisely controlled, as shown in Fig.~\exoref{fig:experiment-concept}{\,a, b}.
A laser beam incident at \SI{6}{\degree} relative to the axis of the TEM column illuminates both the sample and reference, which are traversed by the 200\,keV electron beam focused on the sample to a spot size of about \SI{5}{nm}. The surface normal of the sample plane is tilted by about \SI{13.2}{\degree} from the electron beam, and away from the incident laser beam (Fig.~\exoref{fig:experiment-concept}{\,c}). This ensures that the optical phase fronts propagating at the vacuum speed of light match the modulations of the electron wave function, propagating at the electron group velocity \(v_\text{e}\)$\approx0.695\,c_0$ (Refs.~\citenum{Kirchner2014, Morimoto2018, Kfir2020}). Thus, in the plane of the reference membrane (see Fig.~\exoref{fig:experiment-concept}{\,d}), each part of the weakly conical electron beam interacts with the same optical phase.


As a model system to study nanometric optical excitations~\cite{Nelayah2007}, we use a colloidal triangular gold nanoprism with a thickness of about \SI{7}{nm} and a side length of \SI{100}{nm}. Inelastic electron--light scattering at the optical near field of the nanoparticle induces a PINEM-type electron phase modulation with a spatially dependent coefficient $g$ that is proportional to the local longitudinal electrical field $E_z$ (see Methods)~\cite{GarciadeAbajo2010, Park2010}. In scanning TEM, we raster the focused electron probe across the investigated gold nanoprism, and an electron kinetic energy spectrum is measured at every position to image the spatial distribution of the optical near field.

For a given point in the scanning routine, we observed an interaction-induced sinusoidal modulation of the electron wave with reference phase, as sketched in Fig.~\exoref{fig:experiment-concept}{\,e}. This results from the coherent superposition of light--electron interactions taking place at the sample (coupling coefficient $g$) and at the reference membrane (coupling $g_\text{ref}$). Importantly, both modulations add up coherently~\cite{Echternkamp2016} in the total coupling coefficient $g_\text{total} = g + g_\text{ref}$. The magnitude \(\abs{g_\text{total}}\) of the total phase modulation governs the magnitude and number of populated sidebands and is directly obtained from fits to the measured spectra (see Methods).

As the electron group velocity and the light phase velocity differ, changing the position of the reference plane $\Delta z$ modifies the phase of the reference, which allows for cycling through constructive and destructive interference of both actions ($2\pi$ phase change for $\Delta z=\,$\(\SI{2.1}{\micro m}\)). We record electron spectra in raster scans of the sample plane for a discrete set of reference phases across a complete cycle, selected by displacements of the second membrane. These measurements yield a stack of \(\abs{g_\text{total}}\) maps as a function of the reference phase, as depicted in Fig.~\exoref{fig:experiment-concept}{\,e}.

Recording a complete interferogram (total electron--light coupling strength as a function of relative phase) for every lateral position yields
\begin{equation}\label{eq:g_total}
 \abs{g_\text{total}}^2=\abs{g}^2+\abs{g_\text{ref}}^2+2\, \gamma\, \abs{g} \abs{g_\text{ref}} \cos\left(\varphi_\text{g}-\varphi_\text{ref}\right),
\end{equation}
where $\varphi_\text{g}-\varphi_\text{ref}$ denotes the phase difference between $g$ and $g_\text{ref}$.
A coherence factor $0 \leq \gamma\leq 1$ is introduced here to account for imperfect interference conditions, which arise from amplitude or phase averaging (e.g., due to the finite size of the probe, dispersion effects and deviations from perfect velocity matching at different propagation angles
\footnote{Strictly speaking, dispersive propagation between the interactions planes, leading to attosecond electron bunching \cite{Priebe2017, Morimoto2018, Kozak2018, Ryabov2020}, needs to be taken into account, but this only represents a small correction at the propagation distances and field strengths considered}).

\begin{figure}[tb]
\centering
\includegraphics[]{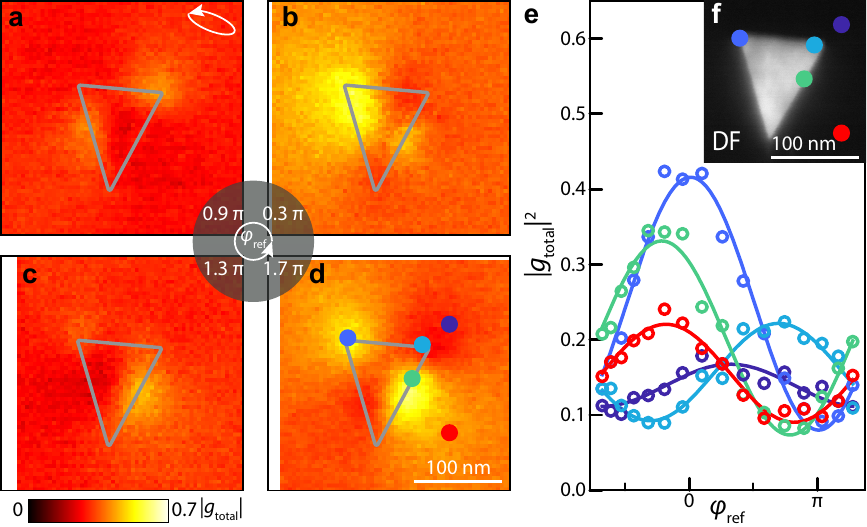}
\caption{\textbf{Recorded FREHD interferograms for an optically excited gold nanoprism.}
\textbf{a-d}, Spatial scans of the magnitude of the total phase modulation for four different reference phases \(\varphi_\text{ref}\).
\textbf{e}, Interferograms (\(|{g_\text{total}}|^2\) 
as a function of relative phase) at the pixel positions indicated by the colour-coordinated dots in \textbf{d} and \textbf{f} (solid lines: sinusoidal fits).
\textbf{f}, Dark-field image of the nanoprism.
}\label{fig:exp-data}
\end{figure}

Figures~\exoref{fig:exp-data}{\,a-d} show images of the measured total interaction strength at four fixed reference phases across the full cycle.
The interference of different plasmonic modes in the nanoprism leads to lobes and nodes at the triangle edge and centre, respectively.
These features are generally aligned to the main axis of the weakly elliptical incident laser polarization (indicated in Fig.~\exoref{fig:exp-data}{\,a}). It is apparent that the intensity of the lobes varies during the interference cycle, and maxima appear at different locations. Figure~\exoref{fig:exp-data}{\,e} displays phase-dependent total coupling strengths for the positions indicated both in Fig.~\exoref{fig:exp-data}{\,d} and in the dark field image in Fig.~\exoref{fig:exp-data}{\,f}. From fits to Eq.~\ref{eq:g_total} at each pixel (solid lines in Fig.~\exoref{fig:exp-data}{\,e}), we find that a uniform magnitude and phase of the reference as well as a single coherence factor \(\gamma=0.68\) describe well the entire data set.

\begin{figure*}[tb]
\centering
\includegraphics[]{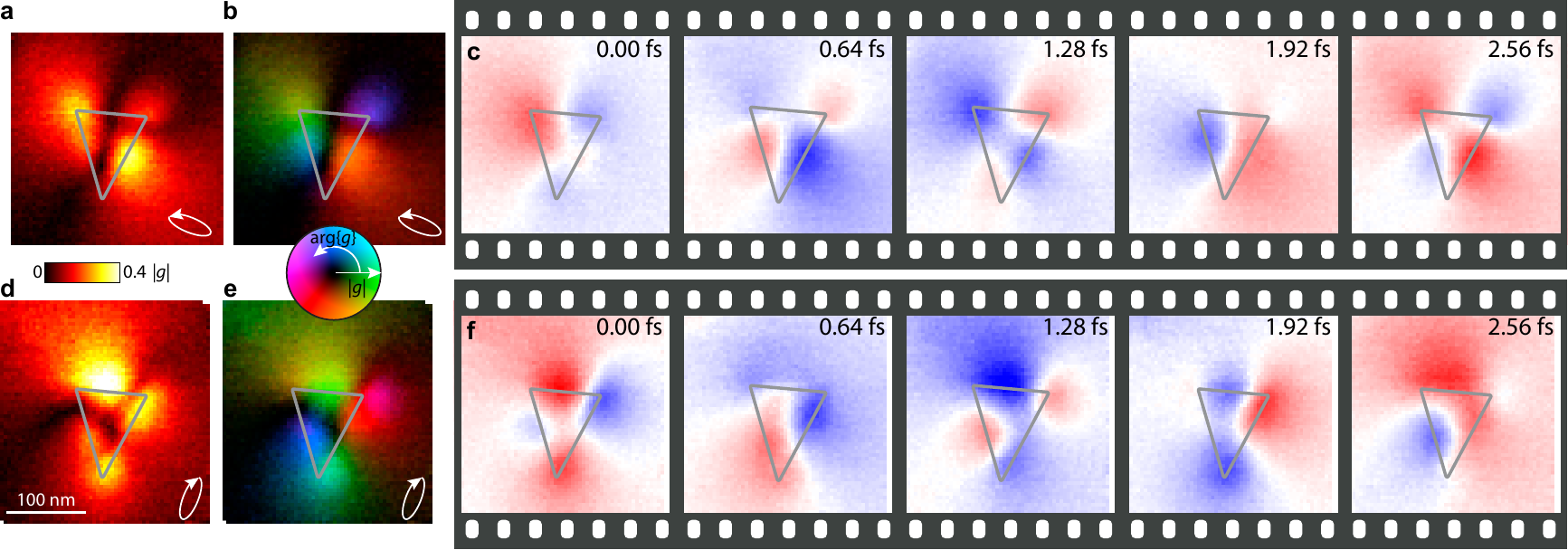}
\caption{
\textbf{Measured electric-field evolution at a gold nanoprism.}
\textbf{a}-\textbf{b},~The interferogram analysis in Fig.~\ref{fig:exp-data} yields the magnitude (\textbf{a}) and complex amplitude (\textbf{b}) of the sample-induced phase modulation $g$. \textbf{c},~Temporal sequence of the out-of-plane electric field evolution within the optical cycle.
\textbf{d}-\textbf{e},~Corresponding measurement for a weakly elliptical polarization state with major axis rotated by \SI{90}{\degree}. The most pronounced maxima in both measurements are clearly aligning with the polarization angle. The complete movies \textbf{c} and \textbf{f} are contained in the Supplementary Information.
}\label{fig:movie}
\end{figure*}

Figure~\ref{fig:movie} displays the complex response of the nanoprism in terms of its magnitude (\lref{fig:movie}{a}) and complex amplitude (\lref{fig:movie}{b}). Here, we subtracted a small uniform background coupling coefficient $g_\text{Si3N4}=0.091$ that stems from the silicon nitride membrane supporting the nanostructure.
The measurement clearly shows that the local maxima of the optical near field exhibit different optical phases. As they represent the complete information about the field, these measurements can be depicted in a time-domain sequence (Fig.~\exoref{fig:movie}{\,c}), illustrating the real part of the out-of-plane optical electric field \(E_z \propto
\abs{g}\cos{(\varphi_\text{g}-\omega t)}\)
as frames of an attosecond movie. Notably, the handedness of the weakly elliptical laser polarization appears as a time-dependent rotation of the maxima around the nanoprism. 
 
We repeat these measurements for a rotation of the major polarization axis by \SI{90}{\degree} (Figs.~\exoref{fig:movie}{\,d-f}), clearly leading to pronounced oscillations along different axes of the nanoprism. The experimental characterization of these complex electric near fields for two non-collinear polarizations constitutes a full polarization-dependent characterization of the near-field optical response. Hence, near fields resulting from arbitrary incident polarization states are immediately retrieved by a corresponding linear combination of these two measurement results. 

In the broader area of phase-resolved near-field imaging using scanning probe techniques~\cite{Zentgraf2008, Schnell2010, Gerber2014, Virmani2021},
photoelectron emission\cite{Kubo2007, Spektor2017, Davis2020}, electron deflection~\cite{Ryabov2016} and Lorentz microscopy of optical fields~\cite{Gaida2022}, as well as the control of quantum emitters~\cite{Heindl2022}, FREHD has a set of unique strengths. It is non-invasive, perfectly linear in its response, allowing for exceptionally high spatial resolution and providing quantitative values for the electrical field. The technique can be applied to recover sub-cycle fundamental and higher-harmonic phase profiles of any local field distribution, including orbital angular momentum states and topological near fields, such as vortices~\cite{Spektor2017}, skyrmions~\cite{Davis2020} and merons~\cite{Dai2020}.

Moreover, the approach is not limited to electromagnetic fields, but will rather detect any modulation imprinted onto an electron beam by an electronic or structural material response. Notably, this covers attosecond charge-density dynamics causing subtle light-driven changes of the structure factor~\cite{Yakovlev2015} at the fundamental frequency and its harmonics. In particular, higher harmonics in the signal response, as well as partial coherence, can be extracted in a straightforward manner using recently established quantum-state reconstruction schemes for free-electron density matrices~\cite{Priebe2017, Shi2020}.

Detecting such small modulations is facilitated by the intrinsic coherent amplification of the signal by the local oscillator, which follows from the linear (rather than quadratic) scaling of the sideband populations with modulation amplitude. For example, a weak modulation $g=0.01$, leading to only a $P_1\approx0.01\%$ sideband population in the absence of a reference, can be amplified to an interferogram with sideband populations varying by $\SI{16.8}{\percent} \pm \SI{0.6}{\percent}$ (\(g_\text{ref} = 0.45\)).
Depending on experimental conditions, such considerable amplifications render the detection of very weak objects feasible by overcoming given levels of background or detector noise. Broadly speaking, FREHD will greatly simplify reaching shot-noise (or quantum) limited detection of small signals. In direct analogy to optical homodyne detection\cite{Yuen1983} or interferometric single biomolecule imaging~\cite{Shintake2008, Shintake2010}, under perfect detection conditions, the technique does not overcome the fundamental signal-to-noise limits of how many electrons are required to quantify the strength of a weakly scattering object (see considerations in Methods). Pixelated event-based detection, as utilized in our work, approaches unity quantum efficiency and features a practical absence of read and dark noise. Nonetheless, even with such detectors, physical backgrounds stemming from, for example, elastic and inelastic scattering cannot be completely avoided. Hence, there is a substantial practical advantage offered by coherent signal amplification, analogously to optical interferometric scattering microscopy~\cite{Ortega-Arroyo2012}. Ultimately, besides the benefit of phase sensitivity, these features may bring individual quantum systems, such as molecules, atoms, or colour centres, closer to spectroscopically enhanced detection in electron microscopy. 

In conclusion, we present a general approach for attosecond phase-resolved electron microscopy at few-nanometer spatial resolution. Quantitative detection of wave function modulations imprinted onto a focused electron probe is achieved using a phase-matched local oscillator interaction at a movable reference plane. Free-Electron Homodyne Detection generalizes the high-resolution measurement of attosecond materials responses in electron microscopy, without a need for electron density bunching, and represents a first realization of a larger class of possible multidimensional optical spectroscopies, which may enable the local readout of couplings, correlations and decoherence.

\bibliography{bibliography}

\begin{thebibliography}{10}
\expandafter\ifx\csname url\endcsname\relax
  \def\url#1{\texttt{#1}}\fi
\expandafter\ifx\csname urlprefix\endcsname\relax\def\urlprefix{URL }\fi
\providecommand{\bibinfo}[2]{#2}
\providecommand{\eprint}[2][]{\url{#2}}

\bibitem{Chapman2011}
\bibinfo{author}{Chapman, H.~N.} \emph{et~al.}
\newblock \bibinfo{title}{Femtosecond {{X-ray}} protein nanocrystallography}.
\newblock \emph{\bibinfo{journal}{Nature}} \textbf{\bibinfo{volume}{470}},
  \bibinfo{pages}{73--77} (\bibinfo{year}{2011}).

\bibitem{Miao2015}
\bibinfo{author}{Miao, J.}, \bibinfo{author}{Ishikawa, T.},
  \bibinfo{author}{Robinson, I.~K.} \& \bibinfo{author}{Murnane, M.~M.}
\newblock \bibinfo{title}{Beyond crystallography: {{Diffractive}} imaging using
  coherent x-ray light sources}.
\newblock \emph{\bibinfo{journal}{Science}} \textbf{\bibinfo{volume}{348}},
  \bibinfo{pages}{530--535} (\bibinfo{year}{2015}).

\bibitem{Zayko2021}
\bibinfo{author}{Zayko, S.} \emph{et~al.}
\newblock \bibinfo{title}{Ultrafast high-harmonic nanoscopy of magnetization
  dynamics}.
\newblock \emph{\bibinfo{journal}{Nat Commun}} \textbf{\bibinfo{volume}{12}},
  \bibinfo{pages}{6337} (\bibinfo{year}{2021}).

\bibitem{Scott2012}
\bibinfo{author}{Scott, M.~C.} \emph{et~al.}
\newblock \bibinfo{title}{Electron tomography at 2.4-\aa ngstr\"om resolution}.
\newblock \emph{\bibinfo{journal}{Nature}} \textbf{\bibinfo{volume}{483}},
  \bibinfo{pages}{444--447} (\bibinfo{year}{2012}).
\newblock \eprint{1108.5350}.

\bibitem{Yang2017}
\bibinfo{author}{Yang, Y.} \emph{et~al.}
\newblock \bibinfo{title}{Deciphering chemical order/disorder and material
  properties at the single-atom level}.
\newblock \emph{\bibinfo{journal}{Nature}} \textbf{\bibinfo{volume}{542}},
  \bibinfo{pages}{75--79} (\bibinfo{year}{2017}).
\newblock \eprint{1607.02051}.

\bibitem{Yip2020}
\bibinfo{author}{Yip, K.~M.}, \bibinfo{author}{Fischer, N.},
  \bibinfo{author}{Paknia, E.}, \bibinfo{author}{Chari, A.} \&
  \bibinfo{author}{Stark, H.}
\newblock \bibinfo{title}{Atomic-resolution protein structure determination by
  cryo-{{EM}}}.
\newblock \emph{\bibinfo{journal}{Nature}} \textbf{\bibinfo{volume}{587}},
  \bibinfo{pages}{157--161} (\bibinfo{year}{2020}).

\bibitem{Corkum2007}
\bibinfo{author}{Corkum, P.~B.} \& \bibinfo{author}{Krausz, F.}
\newblock \bibinfo{title}{Attosecond science}.
\newblock \emph{\bibinfo{journal}{Nature Phys}} \textbf{\bibinfo{volume}{3}},
  \bibinfo{pages}{381--387} (\bibinfo{year}{2007}).

\bibitem{Krausz2009}
\bibinfo{author}{Krausz, F.} \& \bibinfo{author}{Ivanov, M.}
\newblock \bibinfo{title}{Attosecond physics}.
\newblock \emph{\bibinfo{journal}{Rev. Mod. Phys.}}
  \textbf{\bibinfo{volume}{81}}, \bibinfo{pages}{163--234}
  (\bibinfo{year}{2009}).

\bibitem{Calegari2016}
\bibinfo{author}{Calegari, F.}, \bibinfo{author}{Sansone, G.},
  \bibinfo{author}{Stagira, S.}, \bibinfo{author}{Vozzi, C.} \&
  \bibinfo{author}{Nisoli, M.}
\newblock \bibinfo{title}{Advances in attosecond science}.
\newblock \emph{\bibinfo{journal}{J. Phys. B: At. Mol. Opt. Phys.}}
  \textbf{\bibinfo{volume}{49}}, \bibinfo{pages}{062001}
  (\bibinfo{year}{2016}).

\bibitem{King2005}
\bibinfo{author}{King, W.~E.} \emph{et~al.}
\newblock \bibinfo{title}{Ultrafast electron microscopy in materials science,
  biology, and chemistry}.
\newblock \emph{\bibinfo{journal}{Journal of Applied Physics}}
  \textbf{\bibinfo{volume}{97}}, \bibinfo{pages}{111101}
  (\bibinfo{year}{2005}).

\bibitem{Zewail2010}
\bibinfo{author}{Zewail, A.~H.}
\newblock \bibinfo{title}{Four-{{Dimensional Electron Microscopy}}}.
\newblock \emph{\bibinfo{journal}{Science}} \textbf{\bibinfo{volume}{328}},
  \bibinfo{pages}{187--193} (\bibinfo{year}{2010}).

\bibitem{Piazza2013}
\bibinfo{author}{Piazza, L.} \emph{et~al.}
\newblock \bibinfo{title}{Design and implementation of a fs-resolved
  transmission electron microscope based on thermionic gun technology}.
\newblock \emph{\bibinfo{journal}{Chemical Physics}}
  \textbf{\bibinfo{volume}{423}}, \bibinfo{pages}{79--84}
  (\bibinfo{year}{2013}).

\bibitem{Feist2017}
\bibinfo{author}{Feist, A.} \emph{et~al.}
\newblock \bibinfo{title}{Ultrafast transmission electron microscopy using a
  laser-driven field emitter: {{Femtosecond}} resolution with a high coherence
  electron beam}.
\newblock \emph{\bibinfo{journal}{Ultramicroscopy}}
  \textbf{\bibinfo{volume}{176}}, \bibinfo{pages}{63--73}
  (\bibinfo{year}{2017}).

\bibitem{Houdellier2018}
\bibinfo{author}{Houdellier, F.}, \bibinfo{author}{Caruso, G.},
  \bibinfo{author}{Weber, S.}, \bibinfo{author}{Kociak, M.} \&
  \bibinfo{author}{Arbouet, A.}
\newblock \bibinfo{title}{Development of a high brightness ultrafast
  {{Transmission Electron Microscope}} based on a laser-driven cold field
  emission source}.
\newblock \emph{\bibinfo{journal}{Ultramicroscopy}}
  \textbf{\bibinfo{volume}{186}}, \bibinfo{pages}{128--138}
  (\bibinfo{year}{2018}).

\bibitem{Zhu2020}
\bibinfo{author}{Zhu, C.} \emph{et~al.}
\newblock \bibinfo{title}{Development of analytical ultrafast transmission
  electron microscopy based on laser-driven {{Schottky}} field emission}.
\newblock \emph{\bibinfo{journal}{Ultramicroscopy}}
  \textbf{\bibinfo{volume}{209}}, \bibinfo{pages}{112887}
  (\bibinfo{year}{2020}).

\bibitem{Danz2021}
\bibinfo{author}{Danz, T.}, \bibinfo{author}{Domr{\"o}se, T.} \&
  \bibinfo{author}{Ropers, C.}
\newblock \bibinfo{title}{Ultrafast nanoimaging of the order parameter in a
  structural phase transition}.
\newblock \emph{\bibinfo{journal}{Science}} \textbf{\bibinfo{volume}{371}},
  \bibinfo{pages}{371--374} (\bibinfo{year}{2021}).

\bibitem{McKenna2017}
\bibinfo{author}{McKenna, A.~J.}, \bibinfo{author}{Eliason, J.~K.} \&
  \bibinfo{author}{Flannigan, D.~J.}
\newblock \bibinfo{title}{Spatiotemporal {{Evolution}} of {{Coherent Elastic
  Strain Waves}} in a {{Single MoS}} {\textsubscript{2}} {{Flake}}}.
\newblock \emph{\bibinfo{journal}{Nano Lett.}} \textbf{\bibinfo{volume}{17}},
  \bibinfo{pages}{3952--3958} (\bibinfo{year}{2017}).

\bibitem{Feist2018a}
\bibinfo{author}{Feist, A.}, \bibinfo{author}{{Rubiano da Silva}, N.},
  \bibinfo{author}{Liang, W.}, \bibinfo{author}{Ropers, C.} \&
  \bibinfo{author}{Sch{\"a}fer, S.}
\newblock \bibinfo{title}{Nanoscale diffractive probing of strain dynamics in
  ultrafast transmission electron microscopy}.
\newblock \emph{\bibinfo{journal}{Structural Dynamics}}
  \textbf{\bibinfo{volume}{5}}, \bibinfo{pages}{014302} (\bibinfo{year}{2018}).

\bibitem{Kurman2021}
\bibinfo{author}{Kurman, Y.} \emph{et~al.}
\newblock \bibinfo{title}{Spatiotemporal imaging of {{2D}} polariton wave
  packet dynamics using free electrons}.
\newblock \emph{\bibinfo{journal}{Science}} \textbf{\bibinfo{volume}{372}},
  \bibinfo{pages}{1181--1186} (\bibinfo{year}{2021}).

\bibitem{RubianodaSilva2018}
\bibinfo{author}{{Rubiano da Silva}, N.} \emph{et~al.}
\newblock \bibinfo{title}{Nanoscale {{Mapping}} of {{Ultrafast Magnetization
  Dynamics}} with {{Femtosecond Lorentz Microscopy}}}.
\newblock \emph{\bibinfo{journal}{Phys. Rev. X}} \textbf{\bibinfo{volume}{8}},
  \bibinfo{pages}{031052} (\bibinfo{year}{2018}).

\bibitem{Zandi2020}
\bibinfo{author}{Zandi, O.} \emph{et~al.}
\newblock \bibinfo{title}{Transient lensing from a photoemitted electron gas
  imaged by ultrafast electron microscopy}.
\newblock \emph{\bibinfo{journal}{Nat Commun}} \textbf{\bibinfo{volume}{11}},
  \bibinfo{pages}{3001} (\bibinfo{year}{2020}).

\bibitem{Ghimire2019}
\bibinfo{author}{Ghimire, S.} \& \bibinfo{author}{Reis, D.~A.}
\newblock \bibinfo{title}{High-harmonic generation from solids}.
\newblock \emph{\bibinfo{journal}{Nature Phys}} \textbf{\bibinfo{volume}{15}},
  \bibinfo{pages}{10--16} (\bibinfo{year}{2019}).

\bibitem{Dombi2020}
\bibinfo{author}{Dombi, P.} \emph{et~al.}
\newblock \bibinfo{title}{Strong-field nano-optics}.
\newblock \emph{\bibinfo{journal}{Rev. Mod. Phys.}}
  \textbf{\bibinfo{volume}{92}}, \bibinfo{pages}{025003}
  (\bibinfo{year}{2020}).

\bibitem{Herink2012}
\bibinfo{author}{Herink, G.}, \bibinfo{author}{Solli, D.~R.},
  \bibinfo{author}{Gulde, M.} \& \bibinfo{author}{Ropers, C.}
\newblock \bibinfo{title}{Field-driven photoemission from nanostructures
  quenches the quiver motion}.
\newblock \emph{\bibinfo{journal}{Nature}} \textbf{\bibinfo{volume}{483}},
  \bibinfo{pages}{190--193} (\bibinfo{year}{2012}).

\bibitem{Langer2016}
\bibinfo{author}{Langer, F.} \emph{et~al.}
\newblock \bibinfo{title}{Lightwave-driven quasiparticle collisions on a
  subcycle timescale}.
\newblock \emph{\bibinfo{journal}{Nature}} \textbf{\bibinfo{volume}{533}},
  \bibinfo{pages}{225--229} (\bibinfo{year}{2016}).

\bibitem{Higuchi2017}
\bibinfo{author}{Higuchi, T.}, \bibinfo{author}{Heide, C.},
  \bibinfo{author}{Ullmann, K.}, \bibinfo{author}{Weber, H.~B.} \&
  \bibinfo{author}{Hommelhoff, P.}
\newblock \bibinfo{title}{Light-field-driven currents in graphene}.
\newblock \emph{\bibinfo{journal}{Nature}} \textbf{\bibinfo{volume}{550}},
  \bibinfo{pages}{224--228} (\bibinfo{year}{2017}).

\bibitem{Spektor2017}
\bibinfo{author}{Spektor, G.} \emph{et~al.}
\newblock \bibinfo{title}{Revealing the subfemtosecond dynamics of orbital
  angular momentum in nanoplasmonic vortices}.
\newblock \emph{\bibinfo{journal}{Science}} \textbf{\bibinfo{volume}{355}},
  \bibinfo{pages}{1187--1191} (\bibinfo{year}{2017}).

\bibitem{Reimann2018}
\bibinfo{author}{Reimann, J.} \emph{et~al.}
\newblock \bibinfo{title}{Subcycle observation of lightwave-driven {{Dirac}}
  currents in a topological surface band}.
\newblock \emph{\bibinfo{journal}{Nature}} \textbf{\bibinfo{volume}{562}},
  \bibinfo{pages}{396--400} (\bibinfo{year}{2018}).

\bibitem{Cavalieri2007}
\bibinfo{author}{Cavalieri, A.~L.} \emph{et~al.}
\newblock \bibinfo{title}{Attosecond spectroscopy in condensed matter}.
\newblock \emph{\bibinfo{journal}{Nature}} \textbf{\bibinfo{volume}{449}},
  \bibinfo{pages}{1029--1032} (\bibinfo{year}{2007}).

\bibitem{Schultze2010}
\bibinfo{author}{Schultze, M.} \emph{et~al.}
\newblock \bibinfo{title}{Delay in {{Photoemission}}}.
\newblock \emph{\bibinfo{journal}{Science}} \textbf{\bibinfo{volume}{328}},
  \bibinfo{pages}{1658--1662} (\bibinfo{year}{2010}).

\bibitem{Sansone2006}
\bibinfo{author}{Sansone, G.} \emph{et~al.}
\newblock \bibinfo{title}{Isolated {{Single-Cycle Attosecond Pulses}}}.
\newblock \emph{\bibinfo{journal}{Science}} \textbf{\bibinfo{volume}{314}},
  \bibinfo{pages}{443--446} (\bibinfo{year}{2006}).

\bibitem{Antoine1996}
\bibinfo{author}{Antoine, P.}, \bibinfo{author}{L'Huillier, A.} \&
  \bibinfo{author}{Lewenstein, M.}
\newblock \bibinfo{title}{Attosecond {{Pulse Trains Using
  High}}\textendash{{Order Harmonics}}}.
\newblock \emph{\bibinfo{journal}{Phys. Rev. Lett.}}
  \textbf{\bibinfo{volume}{77}}, \bibinfo{pages}{1234--1237}
  (\bibinfo{year}{1996}).

\bibitem{Hentschel2001}
\bibinfo{author}{Hentschel, M.} \emph{et~al.}
\newblock \bibinfo{title}{Attosecond metrology}.
\newblock \emph{\bibinfo{journal}{Nature}} \textbf{\bibinfo{volume}{414}},
  \bibinfo{pages}{509--513} (\bibinfo{year}{2001}).

\bibitem{Barwick2009}
\bibinfo{author}{Barwick, B.}, \bibinfo{author}{Flannigan, D.~J.} \&
  \bibinfo{author}{Zewail, A.~H.}
\newblock \bibinfo{title}{Photon-induced near-field electron microscopy}.
\newblock \emph{\bibinfo{journal}{Nature}} \textbf{\bibinfo{volume}{462}},
  \bibinfo{pages}{902--906} (\bibinfo{year}{2009}).

\bibitem{GarciadeAbajo2010}
\bibinfo{author}{{Garc{\'i}a de Abajo}, F.~J.},
  \bibinfo{author}{{Asenjo-Garcia}, A.} \& \bibinfo{author}{Kociak, M.}
\newblock \bibinfo{title}{Multiphoton {{Absorption}} and {{Emission}} by
  {{Interaction}} of {{Swift Electrons}} with {{Evanescent Light Fields}}}.
\newblock \emph{\bibinfo{journal}{Nano Lett.}} \textbf{\bibinfo{volume}{10}},
  \bibinfo{pages}{1859--1863} (\bibinfo{year}{2010}).

\bibitem{Park2010}
\bibinfo{author}{Park, S.~T.}, \bibinfo{author}{Lin, M.} \&
  \bibinfo{author}{Zewail, A.~H.}
\newblock \bibinfo{title}{Photon-induced near-field electron microscopy
  ({{PINEM}}): Theoretical and experimental}.
\newblock \emph{\bibinfo{journal}{New J. Phys.}} \textbf{\bibinfo{volume}{12}},
  \bibinfo{pages}{123028} (\bibinfo{year}{2010}).

\bibitem{Feist2015}
\bibinfo{author}{Feist, A.} \emph{et~al.}
\newblock \bibinfo{title}{Quantum coherent optical phase modulation in an
  ultrafast transmission electron microscope}.
\newblock \emph{\bibinfo{journal}{Nature}} \textbf{\bibinfo{volume}{521}},
  \bibinfo{pages}{200--203} (\bibinfo{year}{2015}).

\bibitem{Baum2007}
\bibinfo{author}{Baum, P.} \& \bibinfo{author}{Zewail, A.~H.}
\newblock \bibinfo{title}{Attosecond electron pulses for {{4D}} diffraction and
  microscopy}.
\newblock \emph{\bibinfo{journal}{Proc. Natl. Acad. Sci. U.S.A.}}
  \textbf{\bibinfo{volume}{104}}, \bibinfo{pages}{18409--18414}
  (\bibinfo{year}{2007}).

\bibitem{Priebe2017}
\bibinfo{author}{Priebe, K.~E.} \emph{et~al.}
\newblock \bibinfo{title}{Attosecond electron pulse trains and quantum state
  reconstruction in ultrafast transmission electron microscopy}.
\newblock \emph{\bibinfo{journal}{Nat. Photonics}}
  \textbf{\bibinfo{volume}{11}}, \bibinfo{pages}{793--797}
  (\bibinfo{year}{2017}).

\bibitem{Morimoto2018}
\bibinfo{author}{Morimoto, Y.} \& \bibinfo{author}{Baum, P.}
\newblock \bibinfo{title}{Diffraction and microscopy with attosecond electron
  pulse trains}.
\newblock \emph{\bibinfo{journal}{Nat. Phys.}} \textbf{\bibinfo{volume}{14}},
  \bibinfo{pages}{252--256} (\bibinfo{year}{2018}).

\bibitem{Kozak2018}
\bibinfo{author}{Koz{\'a}k, M.}, \bibinfo{author}{Sch{\"o}nenberger, N.} \&
  \bibinfo{author}{Hommelhoff, P.}
\newblock \bibinfo{title}{Ponderomotive {{Generation}} and {{Detection}} of
  {{Attosecond Free-Electron Pulse Trains}}}.
\newblock \emph{\bibinfo{journal}{Phys. Rev. Lett.}}
  \textbf{\bibinfo{volume}{120}}, \bibinfo{pages}{103203}
  (\bibinfo{year}{2018}).

\bibitem{Hassan2018}
\bibinfo{author}{Hassan, M.~T.}
\newblock \bibinfo{title}{Attomicroscopy: From femtosecond to attosecond
  electron microscopy}.
\newblock \emph{\bibinfo{journal}{J. Phys. B: At. Mol. Opt. Phys.}}
  \textbf{\bibinfo{volume}{51}}, \bibinfo{pages}{032005}
  (\bibinfo{year}{2018}).

\bibitem{Ryabov2020}
\bibinfo{author}{Ryabov, A.}, \bibinfo{author}{Thurner, J.~W.},
  \bibinfo{author}{Nabben, D.}, \bibinfo{author}{Tsarev, M.~V.} \&
  \bibinfo{author}{Baum, P.}
\newblock \bibinfo{title}{Attosecond metrology in a continuous-beam
  transmission electron microscope}.
\newblock \emph{\bibinfo{journal}{Sci. Adv.}} \textbf{\bibinfo{volume}{6}},
  \bibinfo{pages}{eabb1393} (\bibinfo{year}{2020}).

\bibitem{Schonenberger2019}
\bibinfo{author}{Sch{\"o}nenberger, N.} \emph{et~al.}
\newblock \bibinfo{title}{Generation and {{Characterization}} of {{Attosecond
  Microbunched Electron Pulse Trains}} via {{Dielectric Laser Acceleration}}}.
\newblock \emph{\bibinfo{journal}{Phys. Rev. Lett.}}
  \textbf{\bibinfo{volume}{123}}, \bibinfo{pages}{264803}
  (\bibinfo{year}{2019}).

\bibitem{Mandel1995}
\bibinfo{author}{Mandel, L.} \& \bibinfo{author}{Wolf, E.}
\newblock \emph{\bibinfo{title}{Optical {{Coherence}} and {{Quantum Optics}}}}
  (\bibinfo{publisher}{{Cambridge University Press}}, \bibinfo{year}{1995}),
  \bibinfo{edition}{first} edn.

\bibitem{Schleich2001}
\bibinfo{author}{Schleich, W.}
\newblock \emph{\bibinfo{title}{Quantum Optics in Phase Space}}
  (\bibinfo{publisher}{{Wiley-VCH}}, \bibinfo{address}{{Berlin ; New York}},
  \bibinfo{year}{2001}), \bibinfo{edition}{1st ed} edn.

\bibitem{Yuen1983}
\bibinfo{author}{Yuen, H.~P.} \& \bibinfo{author}{Chan, V. W.~S.}
\newblock \bibinfo{title}{Noise in homodyne and heterodyne detection}.
\newblock \emph{\bibinfo{journal}{Opt. Lett.}} \textbf{\bibinfo{volume}{8}},
  \bibinfo{pages}{177} (\bibinfo{year}{1983}).

\bibitem{Yakovlev2015}
\bibinfo{author}{Yakovlev, V.~S.}, \bibinfo{author}{Stockman, M.~I.},
  \bibinfo{author}{Krausz, F.} \& \bibinfo{author}{Baum, P.}
\newblock \bibinfo{title}{Atomic-scale diffractive imaging of sub-cycle
  electron dynamics in condensed matter}.
\newblock \emph{\bibinfo{journal}{Sci Rep}} \textbf{\bibinfo{volume}{5}},
  \bibinfo{pages}{14581} (\bibinfo{year}{2015}).

\bibitem{Konecna2020}
\bibinfo{author}{Kone{\v c}n{\'a}, A.}, \bibinfo{author}{Di~Giulio, V.},
  \bibinfo{author}{Mkhitaryan, V.}, \bibinfo{author}{Ropers, C.} \&
  \bibinfo{author}{{Garc{\'i}a de Abajo}, F.~J.}
\newblock \bibinfo{title}{Nanoscale {{Nonlinear Spectroscopy}} with {{Electron
  Beams}}}.
\newblock \emph{\bibinfo{journal}{ACS Photonics}} \textbf{\bibinfo{volume}{7}},
  \bibinfo{pages}{1290--1296} (\bibinfo{year}{2020}).

\bibitem{Yurtsever2012a}
\bibinfo{author}{Yurtsever, A.}, \bibinfo{author}{{van der Veen}, R.~M.} \&
  \bibinfo{author}{Zewail, A.~H.}
\newblock \bibinfo{title}{Subparticle {{Ultrafast Spectrum Imaging}} in {{4D
  Electron Microscopy}}}.
\newblock \emph{\bibinfo{journal}{Science}} \textbf{\bibinfo{volume}{335}},
  \bibinfo{pages}{59--64} (\bibinfo{year}{2012}).

\bibitem{Piazza2015}
\bibinfo{author}{Piazza, L.} \emph{et~al.}
\newblock \bibinfo{title}{Simultaneous observation of the quantization and the
  interference pattern of a plasmonic near-field}.
\newblock \emph{\bibinfo{journal}{Nat. Commun.}} \textbf{\bibinfo{volume}{6}},
  \bibinfo{pages}{6407} (\bibinfo{year}{2015}).

\bibitem{Kfir2020}
\bibinfo{author}{Kfir, O.} \emph{et~al.}
\newblock \bibinfo{title}{Controlling free electrons with optical
  whispering-gallery modes}.
\newblock \emph{\bibinfo{journal}{Nature}} \textbf{\bibinfo{volume}{582}},
  \bibinfo{pages}{46--49} (\bibinfo{year}{2020}).

\bibitem{Henke2021}
\bibinfo{author}{Henke, J.-W.} \emph{et~al.}
\newblock \bibinfo{title}{Integrated photonics enables continuous-beam electron
  phase modulation}.
\newblock \emph{\bibinfo{journal}{Nature}} \textbf{\bibinfo{volume}{600}},
  \bibinfo{pages}{653--658} (\bibinfo{year}{2021}).

\bibitem{Shiloh2022}
\bibinfo{author}{Shiloh, R.}, \bibinfo{author}{Chlouba, T.} \&
  \bibinfo{author}{Hommelhoff, P.}
\newblock \bibinfo{title}{Quantum-{{Coherent Light-Electron Interaction}} in a
  {{Scanning Electron Microscope}}}.
\newblock \emph{\bibinfo{journal}{Phys. Rev. Lett.}}
  \textbf{\bibinfo{volume}{128}}, \bibinfo{pages}{235301}
  (\bibinfo{year}{2022}).

\bibitem{Auad2022}
\bibinfo{author}{Auad, Y.} \emph{et~al.}
\newblock \bibinfo{title}{{{\textmu eV}} electron spectromicroscopy using
  free-space light} (\bibinfo{year}{2022}).
\newblock \eprint{2212.12457}.

\bibitem{Shi2020}
\bibinfo{author}{Shi, C.}, \bibinfo{author}{Ropers, C.} \&
  \bibinfo{author}{Hohage, T.}
\newblock \bibinfo{title}{Density matrix reconstructions in ultrafast
  transmission electron microscopy: Uniqueness, stability, and convergence
  rates}.
\newblock \emph{\bibinfo{journal}{Inverse Problems}}
  \textbf{\bibinfo{volume}{36}}, \bibinfo{pages}{025005}
  (\bibinfo{year}{2020}).

\bibitem{Echternkamp2016}
\bibinfo{author}{Echternkamp, K.~E.}, \bibinfo{author}{Feist, A.},
  \bibinfo{author}{Sch{\"a}fer, S.} \& \bibinfo{author}{Ropers, C.}
\newblock \bibinfo{title}{Ramsey-type phase control of free-electron beams}.
\newblock \emph{\bibinfo{journal}{Nat. Phys.}} \textbf{\bibinfo{volume}{12}},
  \bibinfo{pages}{1000--1004} (\bibinfo{year}{2016}).

\bibitem{Taleb2023}
\bibinfo{author}{Taleb, M.}, \bibinfo{author}{Hentschel, M.},
  \bibinfo{author}{Rossnagel, K.}, \bibinfo{author}{Giessen, H.} \&
  \bibinfo{author}{Talebi, N.}
\newblock \bibinfo{title}{Phase-locked photon\textendash electron interaction
  without a laser}.
\newblock \emph{\bibinfo{journal}{Nat. Phys.}} \bibinfo{pages}{1--8}
  (\bibinfo{year}{2023}).

\bibitem{Madan2019}
\bibinfo{author}{Madan, I.} \emph{et~al.}
\newblock \bibinfo{title}{Holographic imaging of electromagnetic fields via
  electron-light quantum interference}.
\newblock \emph{\bibinfo{journal}{Sci. Adv.}} \textbf{\bibinfo{volume}{5}},
  \bibinfo{pages}{eaav8358} (\bibinfo{year}{2019}).

\bibitem{Note1}
\bibinfo{note}{We note that placing the reference above the sample is of course
  equally possible. Generally, however, dispersive effects must be considered,
  which will differ in both scenarios}.

\bibitem{Kirchner2014}
\bibinfo{author}{Kirchner, F.~O.}, \bibinfo{author}{Gliserin, A.},
  \bibinfo{author}{Krausz, F.} \& \bibinfo{author}{Baum, P.}
\newblock \bibinfo{title}{Laser streaking of free electrons at 25 {{keV}}}.
\newblock \emph{\bibinfo{journal}{Nature Photon}} \textbf{\bibinfo{volume}{8}},
  \bibinfo{pages}{52--57} (\bibinfo{year}{2014}).

\bibitem{Nelayah2007}
\bibinfo{author}{Nelayah, J.} \emph{et~al.}
\newblock \bibinfo{title}{Mapping surface plasmons on a single metallic
  nanoparticle}.
\newblock \emph{\bibinfo{journal}{Nature Phys}} \textbf{\bibinfo{volume}{3}},
  \bibinfo{pages}{348--353} (\bibinfo{year}{2007}).

\bibitem{Note2}
\bibinfo{note}{Strictly speaking, dispersive propagation between the
  interactions planes, leading to attosecond electron bunching \cite
  {Priebe2017, Morimoto2018, Kozak2018, Ryabov2020}, needs to be taken into
  account, but this only represents a small correction at the propagation
  distances and field strengths considered}.

\bibitem{Zentgraf2008}
\bibinfo{author}{Zentgraf, T.} \emph{et~al.}
\newblock \bibinfo{title}{Amplitude- and phase-resolved optical near fields of
  split-ring-resonator-based metamaterials}.
\newblock \emph{\bibinfo{journal}{Opt. Lett.}} \textbf{\bibinfo{volume}{33}},
  \bibinfo{pages}{848} (\bibinfo{year}{2008}).

\bibitem{Schnell2010}
\bibinfo{author}{Schnell, M.} \emph{et~al.}
\newblock \bibinfo{title}{Amplitude- and {{Phase-Resolved Near-Field Mapping}}
  of {{Infrared Antenna Modes}} by {{Transmission-Mode Scattering-Type
  Near-Field Microscopy}}}.
\newblock \emph{\bibinfo{journal}{J. Phys. Chem. C}}
  \textbf{\bibinfo{volume}{114}}, \bibinfo{pages}{7341--7345}
  (\bibinfo{year}{2010}).

\bibitem{Gerber2014}
\bibinfo{author}{Gerber, J.~A.}, \bibinfo{author}{Berweger, S.},
  \bibinfo{author}{O'Callahan, B.~T.} \& \bibinfo{author}{Raschke, M.~B.}
\newblock \bibinfo{title}{Phase-{{Resolved Surface Plasmon Interferometry}} of
  {{Graphene}}}.
\newblock \emph{\bibinfo{journal}{Phys. Rev. Lett.}}
  \textbf{\bibinfo{volume}{113}}, \bibinfo{pages}{055502}
  (\bibinfo{year}{2014}).

\bibitem{Virmani2021}
\bibinfo{author}{Virmani, D.} \emph{et~al.}
\newblock \bibinfo{title}{Amplitude- and {{Phase-Resolved Infrared
  Nanoimaging}} and {{Nanospectroscopy}} of {{Polaritons}} in a {{Liquid
  Environment}}}.
\newblock \emph{\bibinfo{journal}{Nano Lett.}} \textbf{\bibinfo{volume}{21}},
  \bibinfo{pages}{1360--1367} (\bibinfo{year}{2021}).

\bibitem{Kubo2007}
\bibinfo{author}{Kubo, A.}, \bibinfo{author}{Pontius, N.} \&
  \bibinfo{author}{Petek, H.}
\newblock \bibinfo{title}{Femtosecond {{Microscopy}} of {{Surface Plasmon
  Polariton Wave Packet Evolution}} at the {{Silver}}/{{Vacuum Interface}}}.
\newblock \emph{\bibinfo{journal}{Nano Lett.}} \textbf{\bibinfo{volume}{7}},
  \bibinfo{pages}{470--475} (\bibinfo{year}{2007}).

\bibitem{Davis2020}
\bibinfo{author}{Davis, T.~J.} \emph{et~al.}
\newblock \bibinfo{title}{Ultrafast vector imaging of plasmonic skyrmion
  dynamics with deep subwavelength resolution}.
\newblock \emph{\bibinfo{journal}{Science}} \textbf{\bibinfo{volume}{368}}
  (\bibinfo{year}{2020}).

\bibitem{Ryabov2016}
\bibinfo{author}{Ryabov, A.} \& \bibinfo{author}{Baum, P.}
\newblock \bibinfo{title}{Electron microscopy of electromagnetic waveforms}.
\newblock \emph{\bibinfo{journal}{Science}} \textbf{\bibinfo{volume}{353}},
  \bibinfo{pages}{374--377} (\bibinfo{year}{2016}).

\bibitem{Gaida2022}
\bibinfo{author}{Gaida, J.} \emph{et~al.}
\newblock \bibinfo{title}{Lorentz {{Microscopy}} of {{Optical Fields}}}.
\newblock \bibinfo{type}{Preprint}, \bibinfo{institution}{{In Review}}
  (\bibinfo{year}{2022}).

\bibitem{Heindl2022}
\bibinfo{author}{Heindl, M.~B.} \emph{et~al.}
\newblock \bibinfo{title}{Ultrafast imaging of terahertz electric waveforms
  using quantum dots}.
\newblock \emph{\bibinfo{journal}{Light Sci Appl}}
  \textbf{\bibinfo{volume}{11}}, \bibinfo{pages}{5} (\bibinfo{year}{2022}).

\bibitem{Dai2020}
\bibinfo{author}{Dai, Y.} \emph{et~al.}
\newblock \bibinfo{title}{Plasmonic topological quasiparticle on the nanometre
  and femtosecond scales}.
\newblock \emph{\bibinfo{journal}{Nature}} \textbf{\bibinfo{volume}{588}},
  \bibinfo{pages}{616--619} (\bibinfo{year}{2020}).

\bibitem{Shintake2008}
\bibinfo{author}{Shintake, T.}
\newblock \bibinfo{title}{Possibility of single biomolecule imaging with
  coherent amplification of weak scattering x-ray photons}.
\newblock \emph{\bibinfo{journal}{Phys. Rev. E}} \textbf{\bibinfo{volume}{78}},
  \bibinfo{pages}{041906} (\bibinfo{year}{2008}).

\bibitem{Shintake2010}
\bibinfo{author}{Shintake, T.}
\newblock \bibinfo{title}{Erratum: {{Possibility}} of single biomolecule
  imaging with coherent amplification of weak scattering x-ray photons
  [{{Phys}}. {{Rev}}. {{E}} {\textbf{78}} , 041906 (2008)]}.
\newblock \emph{\bibinfo{journal}{Phys. Rev. E}} \textbf{\bibinfo{volume}{81}},
  \bibinfo{pages}{019901} (\bibinfo{year}{2010}).

\bibitem{Ortega-Arroyo2012}
\bibinfo{author}{{Ortega-Arroyo}, J.} \& \bibinfo{author}{Kukura, P.}
\newblock \bibinfo{title}{Interferometric scattering microscopy ({{iSCAT}}):
  New frontiers in ultrafast and ultrasensitive optical microscopy}.
\newblock \emph{\bibinfo{journal}{Phys. Chem. Chem. Phys.}}
  \textbf{\bibinfo{volume}{14}}, \bibinfo{pages}{15625} (\bibinfo{year}{2012}).

\bibitem{Liebig2016}
\bibinfo{author}{Liebig, F.}, \bibinfo{author}{Th{\"u}nemann, A.~F.} \&
  \bibinfo{author}{Koetz, J.}
\newblock \bibinfo{title}{Ostwald {{Ripening Growth Mechanism}} of {{Gold
  Nanotriangles}} in {{Vesicular Template Phases}}}.
\newblock \emph{\bibinfo{journal}{Langmuir}} \textbf{\bibinfo{volume}{32}},
  \bibinfo{pages}{10928--10935} (\bibinfo{year}{2016}).

\bibitem{Bach2019}
\bibinfo{author}{Bach, N.} \emph{et~al.}
\newblock \bibinfo{title}{Coulomb interactions in high-coherence femtosecond
  electron pulses from tip emitters}.
\newblock \emph{\bibinfo{journal}{Structural Dynamics}}
  \textbf{\bibinfo{volume}{6}}, \bibinfo{pages}{014301} (\bibinfo{year}{2019}).

\bibitem{Ciappina2017}
\bibinfo{author}{Ciappina, M.~F.} \emph{et~al.}
\newblock \bibinfo{title}{Attosecond physics at the nanoscale}.
\newblock \emph{\bibinfo{journal}{Reports on Progress in Physics}}
  \textbf{\bibinfo{volume}{80}}, \bibinfo{pages}{054401}
  (\bibinfo{year}{2017}).
\newblock \eprint{1607.01480}.

\bibitem{Sears2008}
\bibinfo{author}{Sears, C. M.~S.} \emph{et~al.}
\newblock \bibinfo{title}{Production and characterization of attosecond
  electron bunch trains}.
\newblock \emph{\bibinfo{journal}{Phys. Rev. ST Accel. Beams}}
  \textbf{\bibinfo{volume}{11}}, \bibinfo{pages}{061301}
  (\bibinfo{year}{2008}).

\bibitem{paper040}
\bibinfo{author}{{Garc\'{\i}a de Abajo}, F.~J.} \& \bibinfo{author}{Howie, A.}
\newblock \bibinfo{title}{Retarded field calculation of electron energy loss in
  inhomogeneous dielectrics}.
\newblock \emph{\bibinfo{journal}{Phys. Rev. B}} \textbf{\bibinfo{volume}{65}},
  \bibinfo{pages}{115418} (\bibinfo{year}{2002}).

\bibitem{Hohenester2018}
\bibinfo{author}{Hohenester, U.}
\newblock \bibinfo{title}{Making simulations with the {{MNPBEM}} toolbox big:
  {{Hierarchical}} matrices and iterative solvers}.
\newblock \emph{\bibinfo{journal}{Computer Physics Communications}}
  \textbf{\bibinfo{volume}{222}}, \bibinfo{pages}{209--228}
  (\bibinfo{year}{2018}).

\bibitem{Lourenco-Martins2023}
\bibinfo{author}{{Louren{\c c}o-Martins}, H.} \emph{et~al.}
\newblock \bibinfo{title}{In preparation} (\bibinfo{year}{2023}).

\bibitem{LiChunming2010}
\bibinfo{author}{{Li, Chunming}}, \bibinfo{author}{{Xu, Chenyang}},
  \bibinfo{author}{{Gui, Changfeng}} \& \bibinfo{author}{Fox, M.~D.}
\newblock \bibinfo{title}{Distance {{Regularized Level Set Evolution}} and
  {{Its Application}} to {{Image Segmentation}}}.
\newblock \emph{\bibinfo{journal}{IEEE Trans. on Image Process.}}
  \textbf{\bibinfo{volume}{19}}, \bibinfo{pages}{3243--3254}
  (\bibinfo{year}{2010}).

\bibitem{JC1972}
\bibinfo{author}{Johnson, P.~B.} \& \bibinfo{author}{Christy, R.~W.}
\newblock \bibinfo{title}{Optical constants of the noble metals}.
\newblock \emph{\bibinfo{journal}{Phys. Rev. B}} \textbf{\bibinfo{volume}{6}},
  \bibinfo{pages}{4370--4379} (\bibinfo{year}{1972}).

\end{thebibliography}
\section*{Methods}
\subsection*{Samples and piezo double holder}
\begin{figure*}[tb]
    \centering
    \includegraphics{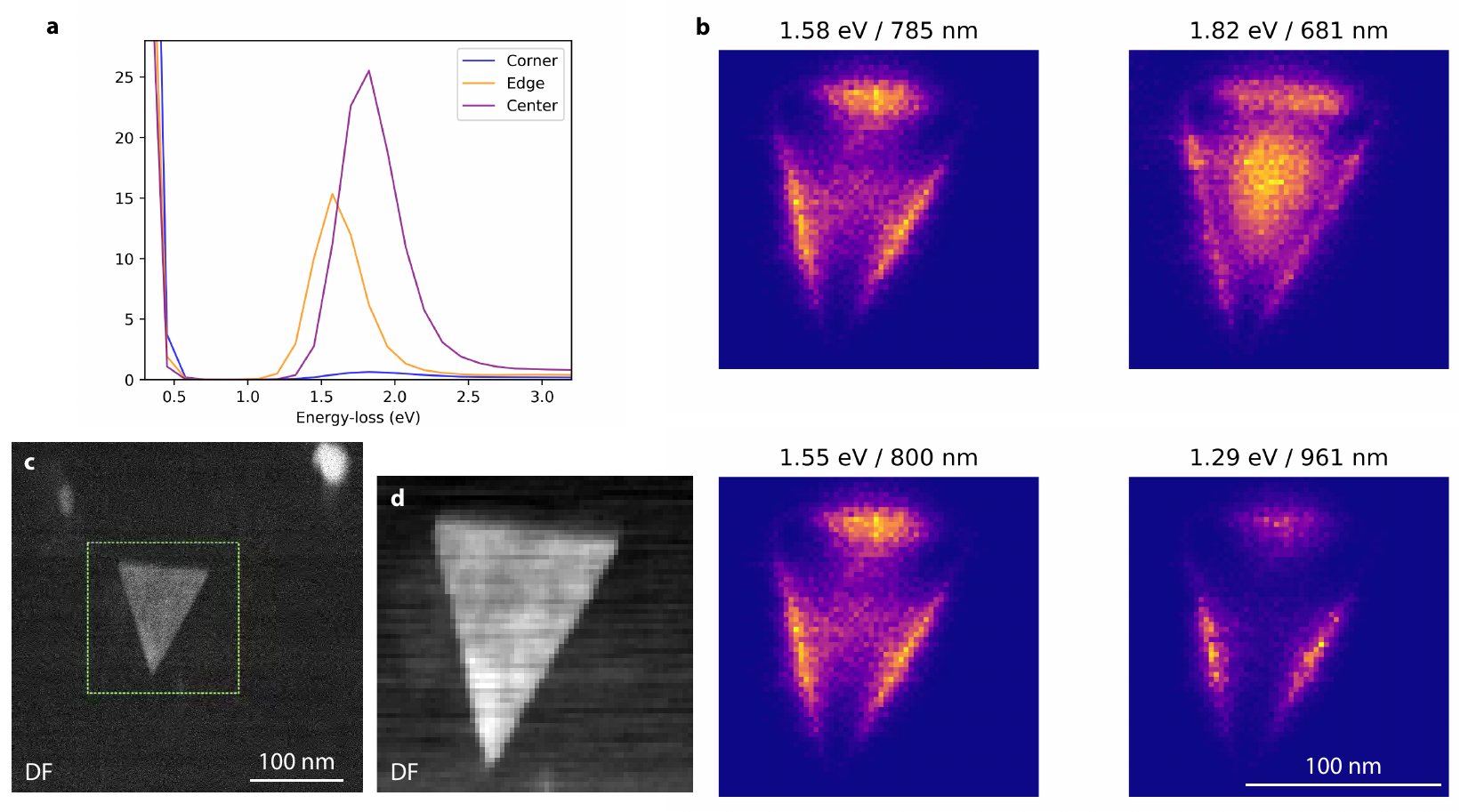}
    \caption{\textbf{Electron energy-loss spectroscopy on a gold nanoprism.}
    \textbf{a} Electron energy-loss spectra taken in standard TEM mode at different positions on the sample showing different energies for different modes (corner, edge and centre).
    \textbf{b} Maps at different energies show the location of different modes.
    \textbf{c},\textbf{d} Dark-field micrographs taken before and during the spectroscopy measurement.
    }\label{fig:sample}
\end{figure*}

The top sample consists of colloidal gold nanoprisms~\cite{Liebig2016} drop cast onto a \SI{50}{nm} thick standard silicon nitride TEM membrane with \SI{100}{\micro\metre} window size. The solution consists of nanoprisms with different geometries, having slightly heterogeneous sizes and shapes with a typical thickness of about \SI{7}{nm}. The nanoprism investigated in our experiments exhibits a side length of about \SI{100}{nm}.
A characterization via electron energy-loss spectroscopy in conventional TEM mode (continuous electron emission) identifies the energy of the edge and centre modes at \SI{1.58}{eV} (\SI{785}{nm}) and \SI{1.82}{eV} (\SI{681}{nm}), respectively. Figure~\ref{fig:sample} shows the corresponding spectra (\lref{fig:sample}{a}) and maps at different energies (\lref{fig:sample}{b}) together with two dark-field (DF) micrographs (\lref{fig:sample}{c,d}). The zero-loss peak has been deconvolved to reveal low energy losses.

The bottom reference sample is a plain standard single crystalline silicon membrane with \SI{30}{nm} thickness and a window size of \SI{500}{\micro m}. It acts like a mirror and produces a homogeneous reference near field. 
The reference sample is positioned \SI{210}{\micro\metre} below the top sample membrane in a custom-designed double holder and can be translated by \SI{2.2}{\micro\metre} relative to the top sample along the membrane's normal (\(z\) direction) by a preloaded, voltage-controlled piezo actuator.

\subsection*{Experimental geometry}
The nanoprism sample and the reference plane are probed simultaneously in scanning TEM mode with \SI{30}{mrad} semi-convergence angle of the electron beam, which is focused down to a \SI{5}{nm} spot diameter on the nanoprism. These focussing conditions lead to an electron beam diameter of about \SI{12.9}{\micro \metre} on the reference membrane. 

The sample and reference planes are both excited with a laser incident at an angle of \SI{-6}{\degree} relative to the electron beam axis. The whole sample holder is tilted along the holder axis by  \SI{13.2}{\degree} from the perpendicular orientation to the electron beam axis (see Fig.~\exoref{fig:experiment-concept}{\,c}) to match the phase velocity of the laser excitation and the group velocity of the \SI{200}{keV} electrons, resulting in a spatially homogeneous optical reference phase along directions normal to the electron beam.

\subsection*{UTEM}
The experiments are carried out at the Göttingen Ultrafast Transmission Electron Microscope (UTEM)~\cite{Feist2017}. Photoelectron pulses are generated from a thermal Schottky field emitter tip using femtosecond laser pulses (centre wavelength \SI{515}{nm}, pulse duration \SI{150}{fs} and \SI{2}{MHz} repetition rate) and accelerated to 200 keV energy.
The electron pulse duration at the sample plane is governed by Coulomb interaction~\cite{Bach2019} leading to about \SI{500}{fs} pulses under the given laser conditions.
Laser excitation of the sample is provided by tunable-wavelength femtosecond pulses from an optical parametric amplifier (here at \SI{960}{nm} central wavelength and \SI{115}{cm^{-1}} bandwidth) and polarization control via retarding quarter and half waveplates. 
In order to provide a time-homogeneous excitation for the duration of the electron pulses, the excitation pulses are stretched to a duration of \SI{1.8}{ps} using a dispersive SF6 glass rod.

The resulting kinetic energy spectrum of the electrons induced by inelastic electron--light scattering is dispersed by a magnetic prism and recorded for each scanning position (pixel dwell time: \SI{30}{ms}) with a hybrid pixel detector. 

The acquisition of \(\SI{64}{px} \times \SI{64}{px}\) with a short dwell time for the recording of the spectrum of \SI{30}{ms} takes roughly \SI{3}{min} per image and \SI{50}{min} in total to record the full interference cycle at 16 different reference phases \(\varphi_\text{ref}\). The resulting dataset consists of \num{65536} spectrograms.

\subsection*{Homodyne detection of sub-cycle light-driven dynamics}

Pump-probe studies of field-driven dynamics are usually performed by combining a harmonic optical excitation (light period $T$) with probing pulses of durations below the optical cycle $T$ (e.g., isolated pulses or pulse trains in the optical domain~\cite{Corkum2007, Ciappina2017}, or recently attosecond electron pulse trains~\cite{Sears2008, Priebe2017, Morimoto2018, Kozak2018}). Sub-cycle light-driven dynamics can also be probed by reconstructing the full quantum state of a system using coherent excitation and probing schemes, thus giving access to its full-time evolution, as performed for interferometric frequency-resolved optical gating (iFROG) or reconstruction algorithms like SQUIRRELS for electrons~\cite{Priebe2017}. This concept is essential for FREHD (Fig.~\ref{fig:homodyne-principle}), which probes complex light-driven dynamics at the nanoscale by coherently modulating the electron wave function both at a sample (coupling $g$) and at a reference (coupling $g_\text{ref}$).

The near-field distribution \(\vec{E}(x,y,z,t) = {\rm Re}\{\vec{E}(x,y,z)e^{-\iu\omega z/v_\text{e}}\}\)  imprints a sinusoidal phase modulation onto the electron wave function as
\[
\Psi(z,t=\infty) = \exp(2\iu\abs{g}\sin(\frac{\omega}{v_\text{e}} z + \varphi)) \Psi(z,t=-\infty).
\]
The interaction is described by the complex coupling parameter
\begin{equation}\label{eq:g_factor}
g(x,y) = \frac{e}{2 \hbar \omega} \int_{-\infty}^{\infty} E_z(x,y,z) e^{-\iu \omega \frac{z}{v_\text{e}}} \text{d}z,
\end{equation}
which represents the spatial Fourier transform of the optical field along the electron beam direction at a spatial frequency \(\Delta k=\frac{\omega}{v_\text{e}}\), where \(v_\text{e}\) is the electron velocity~\cite{GarciadeAbajo2010,Park2010}.
The electron wave function after inelastic electron--light scattering is made up of discrete sidebands labelled by an integer number $N$ with an amplitude given by~\cite{Park2010}
\begin{equation}\label{eq:PsiN_g}
    \Psi_N = J_N(2\abs{g}) e^{\iu N \arg\{-g\}}
\end{equation}
for the sideband of order $N$.
The total transmitted electron probability $P$ is the same of contributions arising from the discrete harmonic sidebands as
\begin{equation}
    P = \sum_{N=-\infty}^\infty P_N,
\end{equation}
where the intensity $P_N = \abs{\Psi_N}^2$ of the sideband $N$ is just the squared modulus of the wave function amplitude.

\begin{figure*}[tb]
\centering
\includegraphics[]{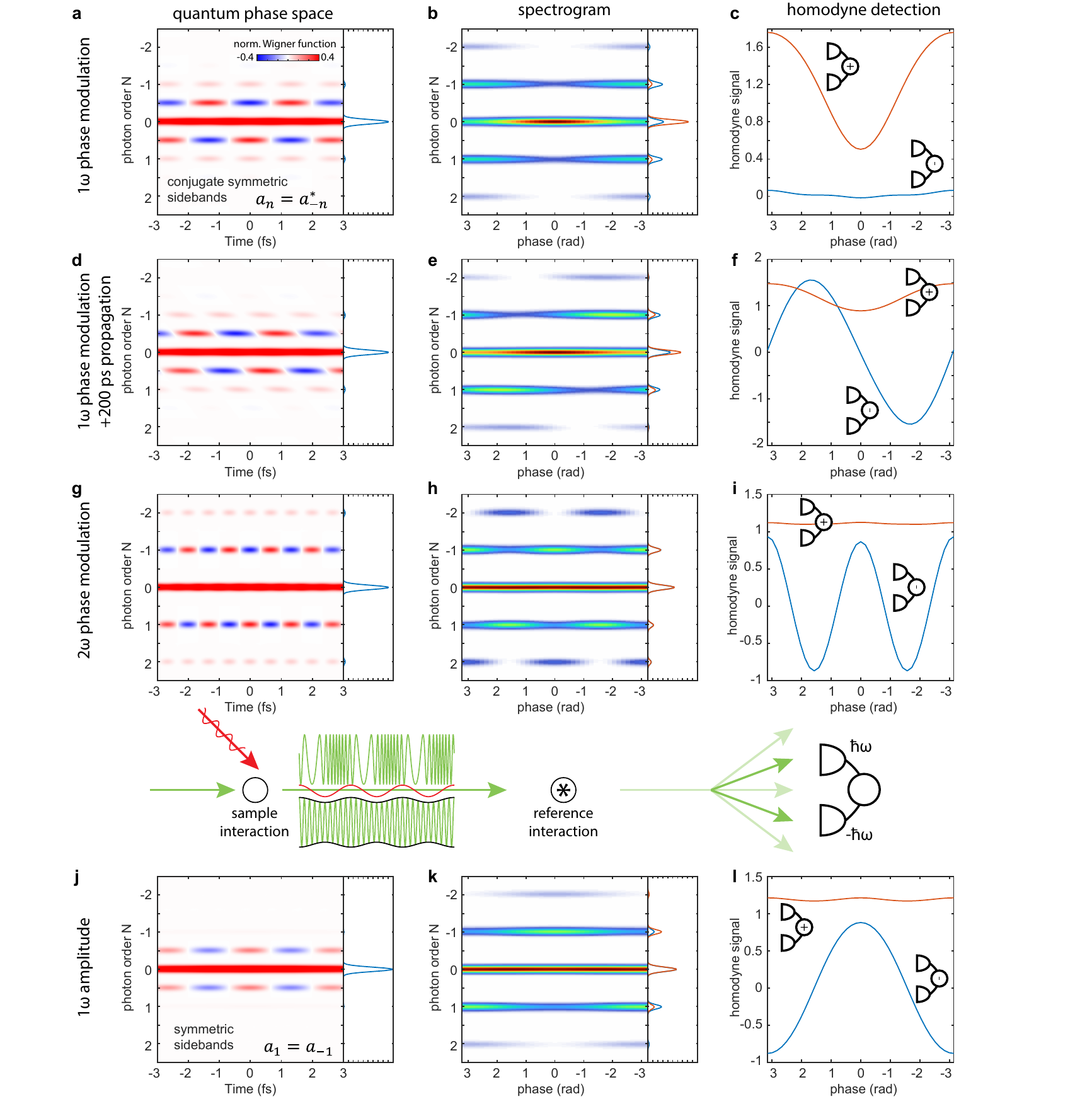}
\caption{
\textbf{Measurement of arbitrary amplitude and phase responses by Free-Electron Homodyne Detection.
} Periodic excitations of an investigated sample resulting in modulations of the amplitude and/or phase of the probing electron wave function at harmonic frequencies $\omega, 2\omega, 3\omega, \dots$ of the exciting optical carrier frequency. These modulations can be probed by the analysis of spectrally symmetric (\textbf{a}-\textbf{f}) and anti-symmetric (\textbf{d}-\textbf{l}) signal channels in FREHD. PINEM-type phase modulation leads to conjugate symmetric and amplitude modulation to symmetric kinetic energy sidebands in the Wigner function (first column). Reference interaction with a pure phase modulation (i.e., convolution with a conjugate symmetric Wigner function; second column) yields either symmetric (\textbf{b}) or anti-symmetric (\textbf{h}, \textbf{k}) spectrograms for non-dispersed electron wavepackets. These spectrograms are evaluated using virtual homodyne detection of the spectral sidebands $P_{-1}$ and $P_{1}$, yielding a phase-dependent signal characteristic for the initial interaction type (third column). Simulation parameters: $g=0.2$, $a=0.2$, $g_{\rm ref}=0.5$, $\lambda = 800$\,nm optical wavelength, $E_0=120$\,keV electron kinetic energy, $t=200$\,ps propagation time. 
}\label{fig:quantum_phasespace}
\end{figure*}

This formalism can be depicted in quantum phase space and generalized to describe amplitude modulations, as shown in Extended Data Fig.~\ref{fig:quantum_phasespace}. Different types of coherent sample interactions lead to phase-dependent signals, either in the symmetric or anti-symmetric detection channels. In quantum phase space, the Wigner distribution $W(x,p)$ describes an electron after optical modulation, with characteristic features arising, such as conjugate symmetric (e.g. phase modulation) or symmetric (e.g. amplitude modulation) sidebands. Mixing a pure phase modulation $W_\text{ref}(x,p)$ as reference results in a phase-dependent spectrogram obtained by convolution of the initial state with the second reference Wigner function: $S(E) = W(x,p) \ast W_\text{ref}(x,p)$. Noteworthy, dispersive propagation between sample interaction and reference leads to a sheering of $W(x,p)$, which transforms an initial phase modulation to an amplitude modulation of the electron wave function, yielding sub-cycle density modulations and attosecond electron pulse trains~\cite{Feist2015}. Besides the possibility of a full quantum state reconstruction using SQUIRRELS\cite{Priebe2017}, FREHD offers fast and direct access to complex harmonic modulations of the electron state by measuring the symmetric $P_1+P_{-1}$ or anti-symmetric $P_1-P_{-1}$ detection channels (weak signal and reference).

\subsection*{Fitting of PINEM spectra}
The spectrum is described by eq.~\ref{eq:PsiN_g}, where each sideband is separated by an energy \(\hbar \omega = \SI{1.29}{eV}\) (the laser photon energy) and convolved with a zero-loss peak of full width at half maximum taken as \SI{0.6}{eV}. 
We fit each recorded spectrum to find the corresponding value of \(\abs{g_\text{total}}\), taking the transmission and a shift of the zero-loss peak as fitting parameters.

\subsection*{Calibration of the reference phase}
\begin{figure*}[tb]
    \centering
    \includegraphics{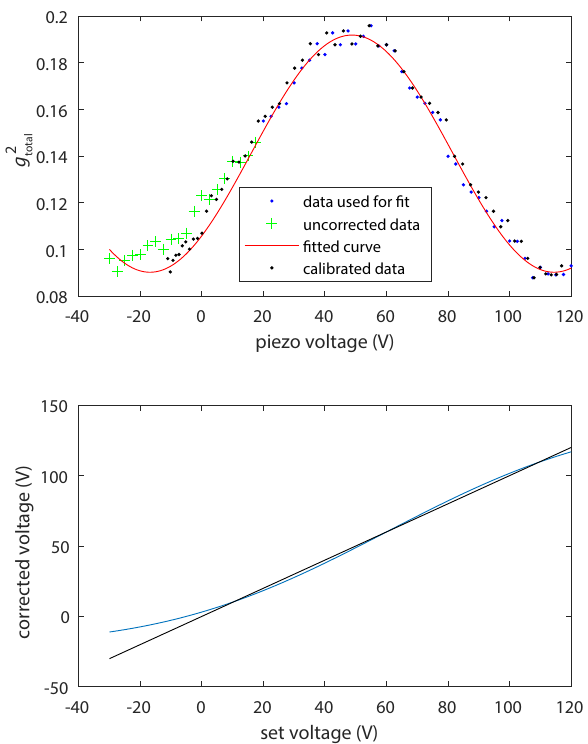}
    \caption{\textbf{Piezo calibration.} 
    The piezo is used both with positive and negative voltages, which leads to a nonlinear dependence of the displacement on the applied voltage, especially when the sign of the voltage changes. We correct the nonlinear movement with an error function, as shown in the bottom plot.
    }\label{fig:piezo}
\end{figure*}

The reference phase is set by changing the \(z\) position of the reference membrane plane with a piezo actuator. The voltage operating the piezo actuator is applied in open-loop mode, requiring a position calibration. 
For positive voltages, the change in position can be assumed to depend linearly on the applied voltage. 
However, to extend the maximum range of motion in this experiment, we also operate the piezo at negative voltages, leading to a nonlinear dependence at the zero-voltage crossing, which is corrected during calibration using an error function in the range of the zero crossing.
We calibrate both the width of the error function and the proportionality of the adjusted voltage to the resulting phase from a measurement with 61 phase steps between 0 and $2\pi$.
This calibration is evaluated in a region clearly separated from the nanoprism on the plain Si$_3$N$_4$ membrane, as shown in Fig.~\ref{fig:piezo}. 
The proportionality and phase offset are determined by fitting a sinusoidal function to the measured interferogram, excluding data points at negative and low voltages (\(<\SI{20}{V}\)). 
The width of the error function is determined such that the measured values at negative and low voltages lie on the extrapolated sinusoidal fit.

\subsection*{BEM simulation} 
\begin{figure*}[tb]
    \centering
    \includegraphics{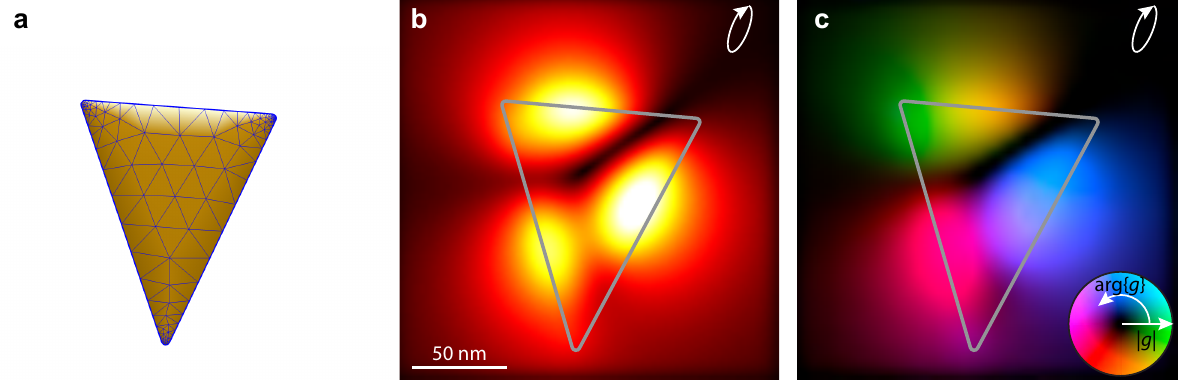}
    \caption{\textbf{BEM electrodynamics simulation.} \textbf{a}, Top view of the surface mesh used in the BEM simulations. The nanoprism shape is extracted from the dark-field image shown in Fig.~\exoref{fig:sample}{\,c} using a DRLSE algorithm~\cite{LiChunming2010}. \textbf{b}, Magnitude and \textbf{c}, phase of the near-field coupling coefficient $g$ simulated with the experimental parameters of Fig.~\exoref{fig:movie}{\,d,e}.
    }\label{fig:simulation}
\end{figure*}

Electrodynamic simulations of the coupling coefficient $g$ are performed using a boundary element method~\cite{paper040} (BEM) as implemented in the \texttt{MNPBEM17} Matlab toolbox~\cite{Hohenester2018}. For a metallic nanostructure of volume $V$ delimited by a boundary $\partial V$ and illuminated by a laser, the coupling coefficient in eq. \eqref{eq:g_factor} can be written in terms of surface boundary sources as~\cite{Lourenco-Martins2023}
\begin{equation}
    g(\vec{R},\omega)=\frac{e}{2\hbar\omega} \oint_{\partial V} \text{d}^2\vec{s}\;\phi(\vec{s}) \Big[ \frac{\omega}{c} h_z(\vec{s})-q_z\sigma(\vec{s}) \Big],
    \label{eq:BEM}
\end{equation}
where $c$ is the speed of light in vacuum, $\phi$ is an electron-generated scalar potential-like function, $\sigma$ and $\vec{h}$ denote the charge and current densities induced at the boundary of the nanostructure, respectively, as obtained from BEM, and the electron velocity is taken along $z$. The nanoprism shape is extracted from experimental dark-field images using a distance regularized level set evolution algorithm (DRLSE, \cite{LiChunming2010}) that renders the 3D meshed surface shown on Fig.~\exoref{fig:simulation}{\,a}. We neglect the presence of the substrate and model the metal response through its frequency-dependent permittivity taken from optical data~\cite{JC1972}. The structure is illuminated by a light plane wave with an incidence angle of \SI{20}{\degree} with respect to the triangle normal, an amplitude of 0.08 V\,nm$^{-1}$ and a wavelength of \SI{560}{nm}. This wavelength differs from the one used in experiment because the absence of the substrate in the simulations causes a blue shift of the plasmonic resonances. The solution of the full retarded Maxwell equations plugged in \eqref{eq:BEM} leads to the maps shown in Fig.~\exoref{fig:simulation}{\,b} and \lref{fig:simulation}{c}.

\subsection*{Shot-noise-limited detection of weak scatterers using FREHD}
Consider an electron scattering channel (e.g., a PINEM sideband) characterised by a small amplitude $g_{\rm total}$ and a probability $|g_{\rm total}|^2\ll1$. For $N$ incident electrons, the expectation value of the number of counts in that channel is $P=N|g_{\rm total}|^2$ with a standard deviation $\Delta P=\sqrt{N}|g_{\rm total}|$ associated with the corresponding Poissonian distribution. In a homodyne measurement approach, $g_{\rm total}=g+g_\text{ref}$ is the sum of an unknown specimen amplitude $g$ and a known reference $g_\text{ref}$ (added, for example, through a second coherent PINEM interaction, as considered in this work). In an experiment, we determine $g$ from the measured $P$ as $g=\sqrt{P/N}-g_\text{ref}$ (assuming $g$ and $g_\text{ref}$ to be real and positive for simplicity), which is measured with an uncertainty $\Delta g\approx\Delta P\big/2\sqrt{NP}$. To determine $g$ with a precision $\eta$, we set $\Delta g=\eta g$, which, using the equations above, leads to a number of incident electrons
\begin{align}
N=\frac{1}{4\eta^2g^2}.
\label{eqforN}
\end{align}
This amounts to $N\approx2500$ required electrons for identifying a weak scatterer with $g=0.1$ (i.e., 1\% scattering probability into the first sideband) at 10\% precision (i.e., $\eta=0.1$). As $N$ is independent of $g_\text{ref}$, it is evident that the homodyne approach does not reduce the number of electrons required to detect the weak signal under perfect shot-noise-limited conditions. Nonetheless, the above-mentioned substantial advantages under real conditions remain.

\section*{Acknowledgements} 
We acknowledge continuing support by the Göttingen UTEM team. We thank R.~M. Sarhan, F.~Liebig and M.\ Bargheer (Institut für Physik und Astronomie, Universität Potsdam) for providing the sample and T. Hohage (Institut für Numerische und Angewandte Mathematik, University of Göttingen) and S. Zayko and S. V. Yalunin for insightful discussions.
\textbf{Funding:}
The experiments were conducted at the Göttingen UTEM Lab, funded by the Deutsche Forschungsgemeinschaft (DFG, German Research Foundation) through Project-ID 432680300 -- SFB 1456 and the Gottfried Wilhelm Leibniz program (RO 3936/4-1), and the European Union’s Horizon 2020 research and innovation programme under grant agreement No. 101017720 (FET-Proactive EBEAM).

\section*{Author contributions}
J.H.G.\ designed and conducted the experiments with support from H.L.-M., M.S.\ and T.R. J.H.G.\ analysed the data with contributions from C.R. H.L.-M.\ performed the BEM simulation. C.R.\ conceived and directed the study, with contributions from F.J.G.d.A.\ and A.F. C.R.\ and J.H.G.\ wrote the manuscript with input from all authors, and all authors discussed the results and their interpretation.

\section*{Competing interests}

The authors declare no competing interests.
\clearpage
\onecolumngrid

\section*{Extended Data}
\FloatBarrier


\end{document}